\documentstyle[12pt,epsf]{article}

\newlength{\pubnumber} 

\catcode`\@=11
\@addtoreset{equation}{section}

\def\section{\@startsection{section}{1}{\z@}{3.5ex plus 1ex minus .2ex}
 {2.3ex plus .2ex}{\large\bf}}
\def\subsection{\@startsection{subsection}{2}{\z@}{2.3ex plus .2ex}
 {2.3ex plus .2ex}{\bf}}

\begin{document}

\begin{titlepage}
\samepage{
\setcounter{page}{1}
\rightline{ACT-8/00}
\rightline{CTP-TAMU-18/00}
\rightline{\tt quant-ph/0007088n}
\rightline{June 2000}
\vfill
\begin{center}
{\Large \bf Quantum Brain?}
\vfill
\vskip .4truecm
\vfill {\large
        Andreas Mershin,$^{1}$\footnote{mershin@rainbow.physics.tamu.edu}
        Dimitri V. Nanopoulos$^{1,2,3}$\footnote{dimitri@soda.physics.tamu.edu}
        and Efthimios M.C. Skoulakis$^{4}$\footnote{eskoulakis@bio.tamu.edu}}
\\
\vspace{.12in}
{\it $^{1}$ Center for Theoretical Physics,
            Dept.\  of Physics, Texas A\&M University,\\
            College Station, TX 77843-4242, USA\\}
\vspace{.06in}
{\it $^{2}$ Astro Particle Physics Group,
            Houston Advanced Research Center (HARC),\\
            The Mitchell Campus,
            Woodlands, TX 77381, USA\\}
\vspace{.06in}
{\it $^{3}$  Academy of Athens, Chair of Theoretical Physics, 
            Division of Natural Sciences,\\
            28 Panepistimiou Avenue, Athens 10679, Greece\\}\vspace{.06in}
{\it $^{4}$ Center for Advanced Invertebrate Molecular Science, 
            Dept.\ of Biology, Texas A\&M University, \\
            College Station, TX 77843-2475, USA \\}

\vspace{.025in}
\end{center}
\vfill
\begin{abstract}
In order to create a novel model of memory and brain function, we focus our 
approach on the sub-molecular (electron), molecular (tubulin) and macromolecular 
(microtubule) components of the neural cytoskeleton. Due to their size and geometry, 
these systems may be approached using the principles of quantum physics.  
We identify quantum-physics derived mechanisms conceivably underlying the 
integrated yet differentiated aspects of memory encoding/recall as well as the molecular 
basis of the engram. We treat the tubulin molecule as the fundamental computation unit 
(qubit) in a quantum-computational network that consists of microtubules (MTs), 
networks of MTs and ultimately entire neurons and neural networks. 

We derive experimentally testable predictions of our quantum brain hypothesis 
and perform experiments on these.
\end{abstract}
\smallskip}
\end{titlepage}

\setcounter{footnote}{0}

% ========================= DEFINITIONS ===================================
%========================================================================
%math definitions
\def\MH#1#2{
$\left( \begin{array}{c}
%row 1 
{#1} \\
%row 2
{#2}
\end{array} \right)$}

\def\MF#1#2#3#4{
$\left( \begin{array}{cc}
%row 1 
{#1}&{#2} \\
%row 2
{#3}&{#4}
\end{array} \right)$}

%Andreas' bra def
\def\bra#1{
$\langle{#1}\left| \right .$}

%Andreas' ket def
\def\ket#1{
$\left|{#1} \rangle \right . $}

%Andreas' new kket def
\def\kket#1{
\left|{#1} \rangle \right .}

%========================================================================
% latex commands
\def\at{ }
\def\beq{\begin{equation}}
\def\eeq{\end{equation}}
\def\beqn{\begin{eqnarray}}
\def\eeqn{\end{eqnarray}}
\def\no{\noindent }
\def\nolabel{\nonumber }

\def\gsim{{\buildrel >\over \sim}}
\def\lsim{{\buildrel <\over \sim}}

% abbreviations
\def\ie{i.e., }
\def\eg{{\it e.g.}}
\def\eq#1{eq.\ (\ref{#1})}

\def\lt{<}

\def\slash#1{#1\hskip-6pt/\hskip6pt}
\def\slk{\slash{k}}

\def\fhalf{\frac{1}{2}}
\def\fsqrt{\frac{1}{\sqrt{2}}}

\def\fsqrtot{\frac{1}{\sqrt{10}}}
\def\fsqrtthreet{\frac{3}{\sqrt{10}}}
\def\fsqrtof{\frac{1}{\sqrt{5}}}
\def\fsqrttf{\frac{2}{\sqrt{5}}}
\def\fsqrtten{\frac{1}{\sqrt{3}}}

\def\half{{\textstyle{1\over 2}}}
\def\third{{\textstyle {1\over3}}}
\def\quarter{{\textstyle {1\over4}}}
\def\sixth{{\textstyle {1\over6}}}
\def\phm{$\phantom{-}$}
\def\m{$\phantom{-}$}
\def\j{$-$}
\def\ps{{\tt +}}
\def\pps{\phantom{+}}

%============================== SECTION 1 ============================

\section{INTRODUCTION}
\subsection{Overview of the Field}
During the last decade or so, it has become increasingly popular among 
researchers to look for manifestations of quantum physics in neurobiological processes 
associated with brain function. Recent works in this field by Penrose\protect\cite{pen1,pen2}, 
Hameroff\protect\cite{ham1}, 
Mavromatos and Nanopoulos\protect\cite{nan1,mn1} and others\protect\cite{sat1} as well as earlier research (as early as 1968) 
on coherent excitations by Fr\" olich\protect\cite{fro1,fro2,fro3}, have been seminal to this new approach to brain 
function research. The arguments for the necessity of this unconventional approach 
have been greatly elaborated upon in the literature by its advocates and yet the very 
existence of "quantum brain" effects is still challenged by physicists and biologists 
alike. To date, at least to our knowledge, experiments targeted at investigating the 
existence of neurobiological quantum phenomena have not been performed. Most of the 
research has been of theoretical and computational nature\protect\cite{mn2,mn3,teg1}and as a result, there has 
been no clear answer. The nature of the subject under investigation is interdisciplinary 
and consequently the target audience has widely varying scientific backgrounds and 
expectations.  The effort described here has included research by experts in both fields 
and aspires to provide a bridge for experimental and theoretical scientists from both disciplines.

Understanding memory will bring us one step closer to finding out how the 
external world is coded in the microscopic structure of the brain and eventually, we will 
be able to appreciate how unique experiences make unique individuals even though the 
basic genetic, molecular and physical processes are shared by all.

\subsection{Problems} 
There are certain aspects of brain function that appear to have no obvious 
explanation based on traditional neuroscience. There exist many biological models of 
memory function but all call for some sort of "Differentiated Yet Integrated"\protect\cite{edel1} (DYI) 
function. Anatomical and neurobiological evidence clearly shows that specific memories 
are not precisely localized in the brain. Although certain structures such as the 
hippocampus\protect\cite{edel1, vor1,haas1} have traditionally been implicated in memory formation more than 
others, it is clear that individual components (for instance correlated visual and auditory 
memories) are stored at macroscopically separated regions of the neural network. This is 
the "differentiated" part of memory. During recall, large numbers of neurons fire in 
tandem to produce an "integrated" picture. By extension, we expect that during the initial 
recording of a memory, a process which results in the {\it engram}, there must also have been 
correlations between distant neurons. This lies at the root of the {\it binding problem} where a 
single stimulus activates neurons located far apart from each other "simultaneously" or at 
least faster than chemical neurotransmission allows. To date, what {\it all} proposed biological 
memory models lack in common, is a plausible mechanism for establishing these fast 
correlations between distant neurons and explaining the {\it speed} at which information is 
processed. This is a feeling shared at least by some biologists who have started looking 
for non-neurotransmitter based communication pathways, such as electrical\protect\cite{der1} and phase 
couplings\protect\cite{fit1}. It is our purpose here to suggest another, quantum physics-derived 
mechanism for the DYI operation of memory. The property of {\it non-locality} exhibited by 
certain quantum systems may produce a solution to the binding problem as well as to the 
speed problem.

Learning and memory are manifested as modifications of behavior produced by 
experience of environmental stimuli and they reflect the function of the brain. Although it 
is generally accepted that changes in the biochemical properties of neurons (especially 
their synapses) mediate changes in brain function and memory encoding, we have yet to 
have a satisfactory understanding of how molecular events effect or influence these 
changes. This is the {\it molecular engram} problem. A prediction of our quantum approach 
gives the neural {\it cytoskeleton} and its associated proteins a major role during engram 
formation and thus proposes experimentally testable molecular mechanisms of memory 
formation.

Classical approaches to digitally simulating biological neural networks (each 
neuron roughly playing the role of a switch whose connections/synapses to other neurons 
are "weighted" according to past experiences) have so far proved insufficient to 
adequately explain how the biological efficiency of recall occurs as well as the observed 
complexity, capacity and versatility of a biological brain. On the other hand, new 
developments in theoretical quantum computation, learning, storage and retrieval 
algorithms, have shown that by using quantum bits or qubits, one resolves the capacity 
problem of classical computers as well as speeds up these processes\protect\cite{ven1}. By modeling the 
brain as a quantum computer we envision to resolve the problems of recall, complexity, 
capacity and versatility.

Assuming our suggestion that quantum phenomena underlie biological function is 
correct, it is yet unclear at exactly which level the transition to classical, purely biological 
processes takes place. With virtually no experimental data in this field, it is impossible to 
precisely define the model but certain testable predictions can nevertheless be derived.

Our quantum mechanical model of brain function differs significantly from the 
classical approach to conventional neural networks but it is not in competition with the 
well established neurobiology of chemical and electrical neurotransmission, synaptic 
function etc. The main difference is that in our model, a single neuron is upgraded from a 
relatively simple (yet adjustable) switch to a device capable of information processing. In 
addition, within the context of our model, (at least some) neurons are capable of launching 
fast connections {\it to establish correlations with distant neurons using the principles of 
quantum entanglement and/or photon interactions}  (both discussed later).

\subsection{Why Use Quantum Mechanics?}
The connection between quantum physical events and biological function has 
been studied for quite some time, for instance Fr\" olich's\protect\cite{fro1,fro2,fro3}work on protein 
conformational changes linked to quantum level interactions/events such as dipole 
oscillation and electron mobility in a protein's hydrophobic pocket. As discussed in great 
detail in references \protect\cite{fro2} and \protect\cite{ham2}, electron density localization inside a hydrophobic pocket 
dictates protein conformation. This should come as no surprise as the van der Waals 
forces arising from a change in the electron localization will push/pull against the charged 
parts of the molecule. As such, this process seems of limited quantum-physical interest since 
analytic solutions to Schr\" odinger equation for such many-body systems are extremely 
difficult to obtain. Motivation for work on quantum mechanics and protein 
conformational changes comes from a defining property of quantum systems discussed 
later, namely their ability to be in a {\it superposition of states} i.e. being in two (or more) 
states at once. 

In particular, the tubulin protein, the structural block of microtubules (MTs), has 
the ability to switch ("flip") from one conformation to another as a result of a shift in the 
electron density localization from one resonance orbital to another (figure1). The tubulin 
system has only two possible basis states labeled \ket{a} and \ket{b} according to whether the 
electrons inside the tubulin hydrophobic pocket are localized closer to the $\alpha$
or $\beta$
monomers. These two states are distinguished from each other by a flip in the electric 
dipole moment vector of the tubulin molecule\protect\cite{sat1} by $29^o$.

\begin{figure}
\epsfxsize=5in
\centerline{\epsffile{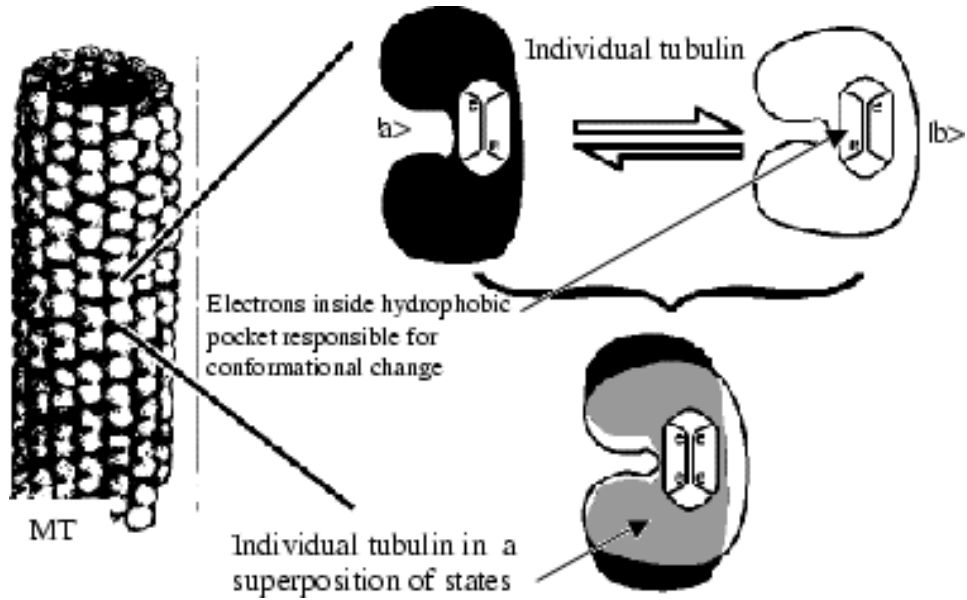}}
\caption{ Relation between microtubule and tubulin. Tubulin can undergo a conformational change from 
the \ket{a} (black) to the \ket{b} (white) basis state depending on the
localization of electrons in its hydrophobic 
pocket. A schematic representation of the superposed state is shown. Modified from Ref. \protect\protect\cite{ham2}.}
\end{figure} 

The tubulin system described above could easily serve as a textbook example of 
how a biological qubit should look like! The two tubulin conformations make for a simple 
binary qubit with the ability of entanglement with similar neighboring qubits/dimers in 
the protofilaments giving us a quantum cluster! The timescale for the spontaneous 
conformational changes in the tubulin dimers is of order  $10^{-11} sec$

Once in an entangled state, a "measurement" or interaction with the environment 
will collapse the state into one of its basis states leaving each tubulin in either the \ket{a} or 
\ket{b} conformations. Yet, the correlations can be communicated instantaneously among the 
tubulin qubits as described in Section 4, spanning entire MTs or conceivably whole 
neurons or neural networks. 

\subsection{Coincidences?}

\vspace{0.3cm}
{\it Alzheimer's Disease}

Damage to neural MTs resulting from hyperphosphorylation of tau ($\tau$) which is a 
microtubule associated protein (MAP), results in memory loss in Alzheimer's Disease 
(AD) patients\protect\cite{vog1} suggesting a connection between MTs and memory. Neurofibrillary 
Tangles (NFTs) are bundles of twisted MTs that are no longer held apart by their MAPs. 
Post-mortem histological examination of AD patients shows a clear and direct correlation 
between NFTs and duration and severity of the disease\protect\cite{arr1}.

\vspace{0.3cm}

{\it Anesthesia}

It is a rather remarkable fact that general anesthesia can be induced by a large 
number of completely different substances of no chemical similarity whatsoever, from 
ether to chloroform to xenon. Purely biophysical studies on the mechanisms of 
anesthesia\protect\cite{fra1,fra2} have shown unequivocally that the long debated action of anesthetics is not 
on the lipid membrane proteins but on the dynamic conformational functions of proteins 
(such as ion channel operation, receptor activation and cytoskeletal function). An 
extension of these findings\protect\cite{ham3} has produced computer simulations strongly suggesting that 
anesthetic molecules bind to the hydrophobic pocket of the tubulin dimer. This is directly 
relevant to our suggestion regarding the role of the tubulin conformational changes as 
follows: binding of an anesthetic molecule to the hydrophobic pocket of the tubulin dimer 
may have the effect of preventing changing the electron orbitals (i.e. the tubulin's ability to 
flip) thus shutting the whole system down. Therefore, in our model, it is just the electric 
dipole properties of these anesthetic substances that need to be similar (which is the case) and not necessarily their chemical properties. Furthermore, if the general anesthetic 
concentrations are not too high, complete reversibility of anesthetic effects is possible, 
indicating that the temporary van der Waals blockage of the crucial tubulin electron(s) has 
ended and conformational changes are free to occur again.

\vspace{0.3cm}

{\it Geometry of Microtubules}

There has been speculation for quite some time that MTs are involved in 
information processing: it has been shown that the particular geometrical arrangement 
(packing) of the tubulin protofilaments obeys an error-correcting mathematical code 
known as the 
$K_{2}(13,2^{6},5)$ code\protect\cite{kor1} (K-code). Error correcting codes are also used in 
classical computers to protect against errors while in quantum computers special error 
correcting algorithms are used to protect against errors by preserving quantum coherence 
among qubits. Furthermore, it has been recently suggested that the geometric curvature of 
MTs may also play a role in information processing\protect\cite{cla1}.

\subsection{Our Motivation}
On the one hand, protein conformational changes are directly related to quantum 
level phenomena and on the other, those same protein functions are directly related to 
system-wide phenomena such as anesthesia and (potentially) memory. Therefore, it seems 
reasonable for us to look for the effects of quantum processes on neuronal (and) brain-
wide function. Lastly, recent theoretical and experimental advances in the field of 
quantum computation call for molecular switches/qubits, the parameters of which fit 
nicely with the proposed role of tubulin dimers. The anticipated quantum clusters also 
sound very much like the MT protofilaments. 

It seems credible that we have uncovered the elementary components of a 
quantum computation network inside the biological brain.

\subsection{Our Research Approach}
Our target system has been the microtubule. We claim that the long and 
characteristically ordered MTs that comprise the bulk of proteins in the axons of 
neurons are the microsites of computation.

During the last few years, physicists have been investigating MTs as physical 
systems applying the principles of Electromagnetic, Quantum and even String 
Theory\protect\cite{nan1, mn1, mn2, mn3}. In the model under discussion here, the MTs' periodic, paracrystalline 
structure, augmented by the K-code, makes them able to support a superposition of 
coherent quantum states among their component tubulin dimers. This quantum 
superposition may collapse spontaneously\protect\cite{nan1,mn1}or dynamically through interactions with the 
environment such as neurotransmitter binding and action potential firing. As a result of 
quantum mechanical entanglement interactions, the MT network in the neuron's axon acts 
in an "orchestrated" or "coherent" way possibly setting up fast communication pathways 
among neurons that do not depend directly on chemical or electrical synaptic signal 
transmission. When the quantum entangled state collapses, the result can be synchronous 
synaptic release of neurotransmitter molecules, and/or feedback information about each 
neurons' environment. The combined effect of such events may be translated into 
orchestrated action and changes in large parts of a neural network. 

Entanglement-based communication would allow MTs to work in tandem and it is 
conceivable that coherence might span macroscopic distances for long times in the brain 
within the context of a particular environment. Although there is a suggestive theoretical 
background\protect\cite{nan1,mn1,mn2,mn3} to justify such assumptions, more experimental 
data is needed before we can say with certainty that quantum coherence is preserved 
for appreciable times and for more than the spatial extent of a few tubulins. 

\subsection{Phenomenology of the Quantum Brain}
As the main areas where we expect to see direct manifestation of quantum 
phenomena are memory encoding, storage and retrieval, these are the points our research 
concentrates upon. If MTs are indeed quantum computing devices, then memory 
encoding would have to be affected by their dynamics. We envision that the role of the 
MAPs, especially MAP-2, is to "tune" the MT network, allowing individual MT states to 
entangle and collapse in specific ways. We expect a redistribution of MAPs to be one of 
the results of memory encoding. We have named this the "guitar string model" (GSM) of 
memory encoding as MAPs can be thought of as the fingers on guitar strings (MTs). 
By changing the binding sites, which in our model represent distinct memory encoding 
events, we change memory encoding (the engram). This is in analogy to different finger 
configurations on guitar strings producing different chords {\it while the strings and fingers 
remain the same}. This model predicts a redistribution of MAP-2 concentration in neurons 
as a result of learning. This has never been conclusively shown and is the goal of our 
experiments described in Section 4. The model also predicts MAP-2 production and 
breakdown as a result of learning and there is some preliminary evidence from other 
groups that this is indeed the case\protect\cite{mar1,woo1}.

\subsection {The Quantum Brain Hypothesis}
What follows is a qualitative description of the proposals of our model in their 
entirety. These are justified later in the text but are included here for completeness. 
\begin{itemize}

\item We propose that the tubulin dimers comprising MTs {\it act as molecular 
binary switches (qubits)} and the two conformations of the tubulin dimer 
are the equivalent of a 0 and 1 in a binary quantum computer. 
\item We propose that information can (at least temporarily) be stored as 
patterns of 0's and 1's corresponding to the conformational states of the 
tubulin dimers. We propose that protofilaments and whole microtubules 
act as memory clusters analogous to RAM (Random Access Memory) in 
digital computers. 

\item We propose that the {\it cytoskeleton} in general and the axonal and dendritic 
microtubules in particular, are the microsites of information manipulation via 
electromagnetic and quantum mechanical interactions between tubulin 
dimers, protofilaments, MTs and MAPs. We further propose that at least 
part of an intermediate or permanent memory trace (the engram) is 
achieved by means of a redistribution of microtubule associated proteins 
along the cytoskeleton. {\it The pattern of MAP binding is the engram}.

\item Following engram formation, neurons that have been simultaneously 
restructured are expected to share similar patterns of MAP binding. 
During recall, neurotransmitter activation of key neuron(s) by a stimulus 
results in instantaneous co-activation of most or all other relevant neurons 
containing a similar cytoskeletal geometry (MAP distribution) via 
quantum coherence phenomena. Thus, large numbers of neurons relevant 
to a particular memory trace can be activated synchronously after which 
ordinary neurotransmitter-based communication sets in.
\end{itemize}

To summarize, these are the problems we will be addressing and the quantum-
physics derived paths to their solution that we propose:
\begin{center}
\begin{tabular}{|c|c|c|}
\hline
{\bf Problem}&
{\bf Relevant quantum property}&
{\bf Research approach}\\
\hline
Binding problem including&Quantum coherence, non-&Theoretical investigation\\
the DYI aspect of memory&locality and entanglement.&yields testable predictions.\\
 &Section 3&Sections 3\&4\\
\hline
Recall&Quantum coherence, non-&Testable predictions\\
 &locality and entanglement.&derived.\\
 &Section 4&Sections 3\&4\\
\hline
Capacity, versatility, speed&Quantum entanglement.&Investigation of quantum-\\
 &Quantum computer-like&computing algorithms\\
 &operation of biological&shown to maintain\\
 &brain.&coherence, increase capacity\\ 
 &Section 4&and speed up recall.\\
 & & Section 4\\
\hline
Molecular basis of the&Quantum effects in neural&Experiments\\
engram&cytoskeletal function.&involving associative\\
 &Tuning of MT network by&learning in fruit-flies are\\
 &MAPs&underway.\\
 &Sections 1--4& Section 5\\
\hline
%\caption{The problems faced by current biological memory models and the novel quantum-physics 
%derived paths proposed towards their solution.}
\end{tabular}
\end{center}

\subsection{Where does our Model Fit in with Classical Neuroscience?}

Existing biological memory models can be complemented by taking advantage 
of the processes suggested in our quantum brain hypothesis. For instance in her 
experiments using rats, Nancy Woolf\protect\cite{woo1,woo2,woo3}, observed degradation of 
MAP-2 in the rat brain 
following an associative learning task. This can be interpreted as follows: MAP-2 
degradation is the first logical step required for the {\it redistribution} of this protein along 
axons and dendrites. Such a redistribution will alter the local geometry of the 
cytoskeleton and this is important for our proposed quantum coherence mechanisms. 
Regrettably, Wolf's analysis was complicated by the multitude and complexity of 
neuronal connections within the mammalian brain, the lack of defined genetic 
background of the animals and the lack of mutants to investigate the mechanism and 
interaction with biochemical pathways known to be operand in learning and memory.

Traditionally, memory is thought to be manifested as Long Term Potentiation 
(LTP). LTP is the process by which a synapse is potentiated meaning that the 
probability of Action Potential (AP) initiation by the follower neuron is increased for 
given excitation of the sending neuron. Our proposed additional role for MAPs in 
memory encoding is not in discord with the LTP hypothesis. It is conceivable that the 
altered dendritic and axonal MT geometry affects synaptic weight and efficacy of signal 
transduction and thus, effects LTP and memory plasticity as an {\it emergent} property, 
necessary for consolidation of memory.

\subsection{Structure of this paper}
In Section 2, we give a brief account of fundamental relevant concepts in 
neurobiology. In Section 3, we offer a simplified, qualitative explanation of the 
formation of coherent quantum modes in MTs. In Section 4, we present an elementary 
introduction to some pertinent concepts of quantum computing and quantum mechanics 
while maintaining the focus on the biological connection. In Section 5, we present our 
experimental design and some preliminary results. Finally, in Section 6 we summarize 
our findings and address discussions of this work by others.

\newpage
\newpage
\section{FUNDAMENTALS OF NEUROBIOLOGY}

\subsection{Cells of the Nervous System}
Neurons are polarized cells which are highly specialized to receive, process, 
transfer and store information. They are subdivided into three major parts. The {\it soma} 
contains the nucleus while the {\it dendrites} are relatively short, multiply branched extensions. 
The single {\it axon} is a long extension that branches at the neuron's distal end. This differential 
architecture reflects functional differences between the two types of projections. 
Information is generally received by the dendrites and transmitted through the soma to 
the axon where it is relayed to the dendrites of neighboring neurons. Multiple neurons 
may be stimulated by one axon and one neuron may receive multiple axonal 
stimulations in each of its dendrites (see figure 2). This complexity is directly correlated 
to the vast capacity of the brain as a whole to receive, process and store information. It seems that there is a signal integration zone at the beginning of the axon known as the "spike initiation zone" (SIZ) indicated by the arrows in figure 2. The SIZ is where action potentials are initiated.

\begin{figure}
\epsfxsize=5in
\centerline{\epsffile{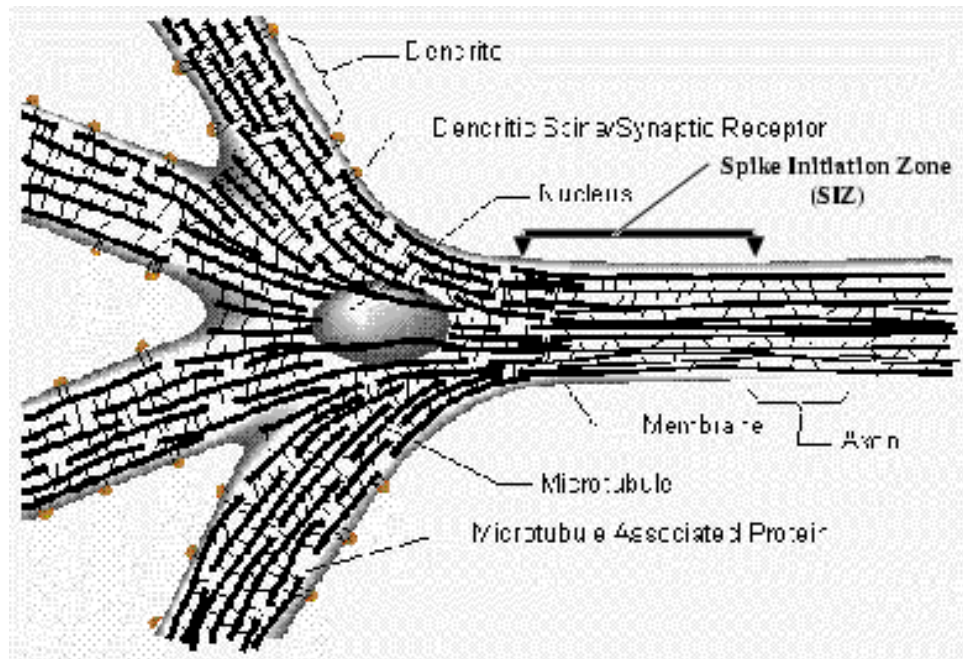}}
\caption{Schematic representation of a neuron. Dendrites, MTs, MAPs and the SIZ are shown. Modified 
from \protect\cite{ham2}}
\end{figure}

\subsection{Neuronal Signaling}
Neurons relay information to each other via specialized structures at the 
dendritic and axonal termini known as synapses. The number of synapses and efficacy 
of information flow through them is a function of the frequency and "importance" of 
information exchanged between these neurons. This neuronal property is plastic in the 
sense that the number of synapses and their efficacy changes as a result of prior and 
frequent use and this reflects memory at the cellular level. These structural changes are 
mediated by biochemical signaling pathways that relay the information flow to various 
areas of the cytoplasm and the nucleus, which in turn responds by altering gene activity 
to mediate the aforementioned events that underlie neuronal plasticity. The evidence 
supporting this classical neurobiological model is overwhelming. However, despite the 
vast number of possible connections within the brain, it is still difficult to explain the 
amount and speed at which information is processed and stored within this tissue based 
purely on amount and strength of synapses available. The hypotheses and proposals 
presented here are not in competition with the well-established neurobiological 
properties of cells. We hope that most if not all observed neurobiological phenomena 
can be explained as {\it emergent} properties of our model. Our proposals extend traditional 
findings by focusing on the generally neglected role of the neuronal (microtubular) 
cytoskeleton and its accessory proteins during information storage and retrieval. 

\subsection{Cytoskeleton and Microtubules}
Neurons as well as all other eukaryotic cells are internally organized and held 
together by a scaffold made up of a network of protein polymers called the cytoskeleton. 
The cytoskeleton consists of microtubules, actin filaments, intermediate filaments and 
microtubule associated proteins (MAPs), which among other functions, link parallel 
arrays of MTs into networks (figure 3). 

\begin{figure}
\epsfxsize=5in
\centerline{\epsffile{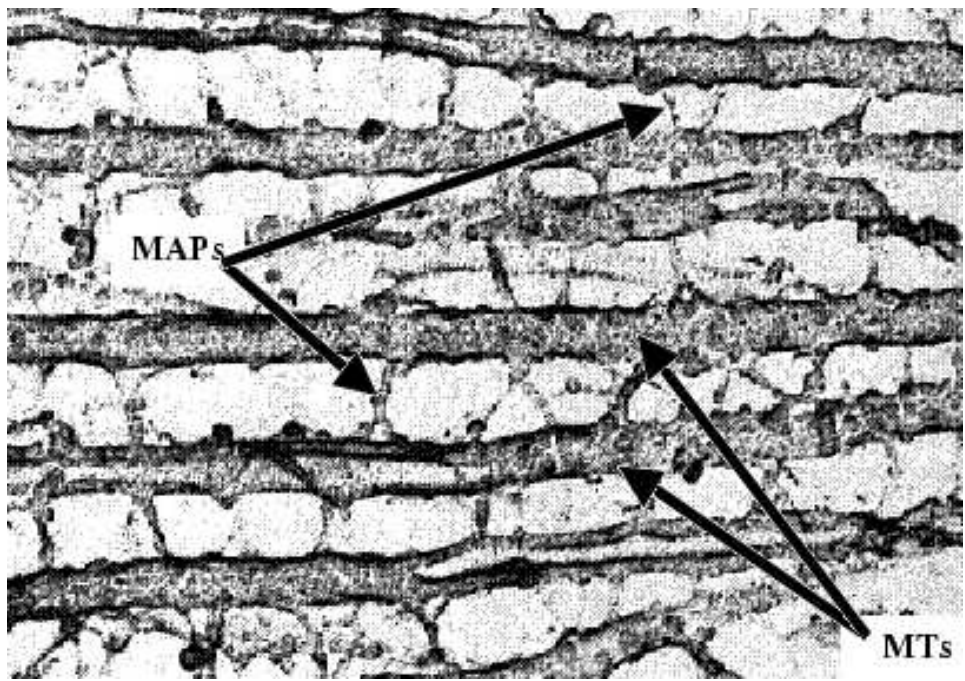}}
\caption{Photograph (micrograph) of flagellar microtubular network. Neural MT networks have similar 
geometry. Modified from \protect\cite{hir1}.}
\end{figure}

The MT's cylindrical walls (outer diameter 25nm, inner diameter 15nm) are 
comprised of 13 longitudinal protofilaments. These protofilaments are constructed from 
a series of subunit proteins known as tubulins (figure 4). Each tubulin subunit is a polar 
dimer of length of about 8nm and it consists of two slightly different classes of a 4nm, 
55kD (kilo-Dalton) monomer known as $\alpha$ and $\beta$-tubulin. The tubulin dimer subunits 
within MTs are arranged in a hexagonal twisted lattice, and helical pathways that repeat 
every 3, 5 and 8 rows (figure 4).

Microtubules are major components of the cell's cytoskeleton and are involved in 
a variety of functions such as mitosis, axonal protein transport, signal transduction and 
--we claim-- quantum computation. These processes are dependent on the distinctive 
structure of the MT. 

\begin{figure}
\epsfxsize=2in
\centerline{\epsffile{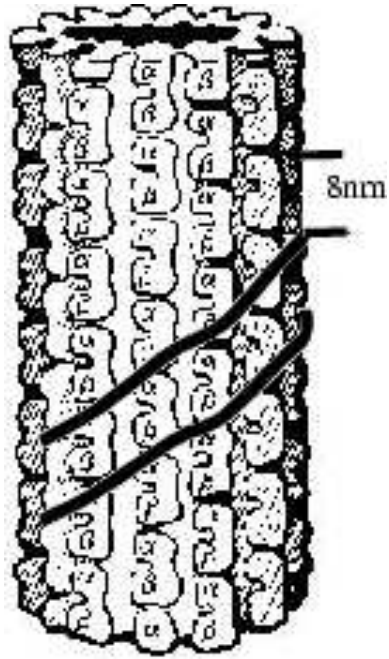}}
\caption{Segment of a microtubule showing tubulin dimers. The structure has been derived using x-ray 
crystallography (Amos and Klug, 1974). Tubulin subunits are 8 nanometer (nm) dimers comprised of 
$\alpha$ and $\beta$ monomers. Modified from Amos \& Klug\protect\cite{amos1}}
\end{figure}

We will concentrate our analysis to axonal MTs of neurons. The axonal MT is 
typically long (hundreds of nm) (figures 2 and 3) and also it is characteristically stable 
(compared to other cytoskeletal MTs which exhibit great dynamic instability). We have 
compelling theoretical indications that MTs are ferroelectric\protect\cite{mn4} and experiments are 
currently underway to confirm our predictions. 

Furthermore, it has been suggested\protect\cite{mn1} that the ordered arrangement of water 
molecules provides isolation from thermal oscillations, and other potential decohering 
mechanisms, thus creating an environment that can support quantum entangled states of 
the component tubulin molecules. This is discussed in more detail in Sections 3 and 4.

\subsection{Microtubule Associated Proteins}
There are many types of MAPs each with different roles in cell function\protect\cite{kan1}. We 
are particularly interested in MAP-2 and MAP-tau, as in our model, MAP-2 
phosphorylation (breakdown) and de-phosphorylation seems to play a major role in 
memory encoding and this has been suggested for some time\protect\cite{fuk1}. MAP-2 consists of a 
pair of high molecular mass (280kD) proteins (isoforms a and b) and a low mass (70kD) 
polypeptide (isoform c). There is experimental evidence that MAP-2c may be 
dephosphorylated following contextual memory training in rodents\protect\cite{fuk1}. Phosphorylation 
of MAP-2 decreases its co-assembly (binding) to MTs\protect\cite{woo1,woo2,woo3,kan1} thus enabling 
cytoskeletal restructuring and favoring (dendrite) plasticity.

\newpage
\section{FORMATION OF COHERENT STATES IN MICROTUBULES}

\subsection{Ordered Water and Superradiance}
There is evidence that the hollow interior of MTs may be capable of supporting 
a very special state of "ordered" water molecules both inside and outside of the 
MT\protect\cite{mn1,mn4}. 
We also notice that it has been recently confirmed experimentally that at the exterior of 
the MT cylinders, there do exist thin layers of charged ions, of thickness of order 7-8 \AA, 
in which the electrostatic interaction energy is larger than the thermal energy due to the 
interaction with the environment\protect\cite{sac1} meaning that electrostatic interaction effects are 
dominant. In view of such results, we have previously conjectured\protect\cite{mn1} that similar layers 
might also exist in the interior of the MT cylinders, which provide us with the 
necessary thermal isolation to sustain quantum coherent states over time scales 
comparable to the dynamical timescales of neural cells, namely of order $10^{-4}-10^{-3} sec$. 
This would make the MT interior act as a waveguide to photons of special frequencies 
and would also thermally isolate the MT interior from the environment, so that it may 
act in a laser-like way, a property called superradiance\protect\cite{del1,del2,jibu1}. Due to the strong 
suppression of such couplings in the disordered states of regular, liquid water, this is not 
ordinarily observed. It is however quite plausible that such behavior characterizes the 
ordered water molecules that exist in the interior of MTs. The presence of unpaired 
electrons in the tubulin molecule is crucial to such a phenomenon. If true, then this 
coupling of the tubulins' electric dipole to the quantum radiation will be responsible for 
the appearance of collective quantum coherent modes\protect\cite{mn4}. Such modes are termed `dipole 
quanta'. This mechanism has been applied to microtubules\protect\cite{jibu1}, with the conclusion that 
such coherent modes cause superradiance, i.e. create a special quantum-mechanical 
ordering of the water molecules with characteristic collapse times much shorter than 
those of thermal interaction and thus make the interior of MTs transparent to photons of 
certain frequencies. This has been conjectured as early as 1978 and MTs have been 
theorized to play the role of `dielectric waveguides' for photons\protect\cite{ham4}. 

Such a coupling implies a `laser-like' behavior. The interaction of the dipole-
quanta coherent modes with the protein dimers results in an entanglement which we 
claim is responsible for the emergence of soliton\footnote{Solitons are special pulse-like 
waveforms that have well-established yet unusual physical properties including non-dispersion 
over large distances of propagation. Solitons are extremely resilient to noise and 
propagate unaltered even after interaction with other solitons.} quantum coherent states, extending 
over large scales, e.g. the MT or even the entire MT network. An explicit mathematical 
construction of such solitonic states has been made in the quantum field-theoretic model 
for MT dynamics of Mavromatos and Nanopoulos\protect\cite{mn2,mn3} in 1997, which was based on 
classical ferroelectric models for the displacement field discussed in other works\protect\cite{sat1}. The 
quantum-mechanical picture described here should be viewed as a simplification of the 
field-theoretic formalism, it is however sufficient for qualitative estimates of emergent 
properties including expected decoherence times. 

\subsection{How is this relevant to information processing?}
To summarize, at least theoretically, there exists a credible mechanism for the 
formation of quantum coherent modes in the water in and around the MT, inspired by 
earlier suggestions of `laser-like' behavior of the water\protect\cite{del1,del2,jibu1} arising from the interaction 
of the electric dipole moments of the water molecules with photons of specific 
frequencies (i.e. selected modes of the quantized elelctromagnetic radiation). This is 
important as it provides the needed isolation from environmental noise to preserve the 
delicate quantum coherence between the qubits/tubulin dimers.

\subsection{Possible Role for the Photons}
This is a variation of the mechanism suggested earlier to establish fast (speed of 
light) connections between distant neurons and integrate information processing inside 
the MT network of a single neuron or a network of neurons. In this scenario, 
entanglement of neural cells over macroscopic distances is not required. Conceivably, 
photons emitted by MTs can be absorbed by distant neurons that are 'tuned' to receive 
at specially modulated frequencies. Again, the tuning/modulation can happen via the 
binding of MAPs which brings us back to the GSM model of memory encoding. It is 
exciting to note that in most of the various suggested quantum computer designs, 
photons are used to communicate information from qubits to detectors.

\subsection{Why no data?}
Whether this is the case in all cell MTs, or only in certain areas, such as the 
characteristically long and stable neural MTs, is something that cannot be determined 
presently. Questions like these can only be answered once detailed information at the 
atomic scale, becomes available on the structure of tubulin dimers, on the precise 
magnitude of their electric dipole moments, and on the detailed structure of the water 
interior to MTs. As a first step in this direction we mention the atomic resolution map 
of tubulin, which became available only recently by means of electron crystallogaphy\protect\cite{nog1}.

\newpage
\section{BASICS OF QUANTUM COMPUTATION}

\subsection{Why Quantum Computation?}
In our quantum brain hypothesis the brain is modeled as a vast network of 
interconnecting neurons which have the potential of isolated and parallel quantum 
computation. As a result, in order to understand this hypothesis it is necessary to grasp the 
basics of quantum computation. 

The possibility of creating quantum computers is being thoroughly investigated by 
numerous research groups around the world both theoretically\protect\cite{shor1,gro1} 
and experimentally\protect\cite{sha1} 
(for a review see Preskill\protect\cite{pre1}). Quantum computation envisions quantum computers 
utilizing qubits rather than conventional bits. Some of the advantages of quantum 
computers of the future are better speed, gigantic memory capacity and immense 
computing power. The most intriguing aspect of quantum computers is their ability to 
perform tasks that are simply impossible using classical computers. The two most 
celebrated such abilities are efficient factorization of large numbers\protect\cite{shor1} and search through 
unordered data lists in times faster than allowable classically\protect\cite{gro1}. Behind both these feats 
lies an integrated and "delocalized" way of handling data that makes the machine capable 
of retrieving stored information in much the same way the human brain does during 
pattern recognition. Furthermore, the existence of a quantum-error correcting code is 
needed to protect the delicate coherent qubits from decoherence. This has been the major 
problem of quantum computers until the works of Shor and Steane have independently 
shown that such a code can be implemented\protect\cite{shor1,ste1}. We conjecture that the 
{\it K-code apparent 
in the packing of the tubulin dimers and protofilaments is partially responsible for keeping 
coherence among the tubulin dimers}. By simulating the brain as a quantum computer it 
seems we are capable of obtaining a more accurate picture than if we simulate the brain as 
a classical, digital computer.

Although based on the well-established physical principles of quantum 
mechanics, quantum computers are yet to be experimentally realized. This has not 
deterred theoretical work in the field and even the writing of "quantum software" in the 
form of mathematical algorithms that take advantage of quantum computers yet to be 
developed\protect\cite{ven1,shor1, gro1}. Note that recently there has been progress towards creating the 
necessary apparatus that will ultimately provide us with quantum logic gates\protect\cite{sha1}

\subsection{Quantum Mechanics and Quantum Computing}

\vspace{0.3cm}
{\it What's so  different about quantum computers?}

The main difference between classical, digital computation (on which the 
traditional approach to simulating the brain using neural networks is based) and quantum 
computation is the latter's usage of quantum bits rather than ordinary bits. 
In our approach, we treat the tubulin dimer as a quantum two-state system which 
represents one qubit i.e. the building block of a quantum cluster. Tubulin protofilaments 
that make up MTs play the role of quantum clusters and whole MTs can be thought of as 
blocks of clusters.

\vspace{0.3cm}
{\it What's so  different about Quantum Mechanics?}

For three quarters of a century, quantum physics has been universally accepted as 
the most accurate account of the phenomena of the microcosm and arguably the most 
accurate and precise scientific theory ever. Atoms, nuclei and elementary 
particles can be correctly described 
only by using the mathematical framework of quantum physics called quantum 
mechanics. The name is an analogy to classical Newtonian mechanics, which can be 
derived as an approximation to quantum mechanics when the objects of study are large in 
mass. Although Newtonian mechanics has been used for centuries to adequately explain the motion 
of everyday macroscopic objects, it proved grossly inadequate with the 
discovery of the atom.

\vspace{0.3cm}
{\it Introduction to Quantum Mechanics.}

The account that follows has been written in the fashion of an introduction to the 
subject, requiring no more mathematical ability than elementary algebra. It is intended to 
give a "first taste" of the quantum mechanical approach and discuss the relevance of 
entangled states to fast communication of correlations among neurons. A complete, fully mathematical 
treatment of MTs can be found in a number of sources\protect\cite{nan1,mn1,mn2,mn3}. Due to the special 
properties exhibited by microscopic systems, special jargon, often counterintuitive to the 
unseasoned reader must be employed. 

Quantum systems have two modes of evolution in time. The first, governed by 
Schr\" odinger's equation (see below) describes the time evolution of quantum systems 
when they are undisturbed by measurements. 'Measurements' are defined as interactions 
of the system with its environment. As long as the system is sufficiently isolated from the 
environment, it follows Schr\" odinger's equation. If an interaction with the environment 
takes place, i.e. a measurement is performed, the system abruptly decoheres i.e. collapses 
or reduces to one of its classically allowed states. 

In what follows we will employ Dirac bracket notation, where the ket \ket{a} is 
analogous to a column vector \MH{a}{b}\ , and the bra \bra{a} 
is the complex conjugate transpose of 
$\left| a \rangle \right . $ which means it is a row vector where all entries have been complex-conjugated.  
$(a^{*},b^{*})$.
Time evolution of quantum systems (in the absence of measurements) is described 
by the Schr\" odinger equation: 
$i\hbar\frac{\partial}{\partial t}\left|\Psi\rangle \right . =H\left|\Psi\rangle \right . $  
where H is the Hamiltonian (energy) 
operator (see below), $i = \sqrt{-1}$ and $\hbar$ is Planck's constant divided by 2$\pi$.

{\it Linear superposition} is a generalization of the familiar mathematical principle of 
linear combination of vectors. Instead of using three orthogonal axes as a basis, quantum 
systems are described by a {\it wavefunction} \ket{Psi}  that exists in a multi-dimensional "Hilbert 
Space"\protect\cite{sak1}. The Hilbert space has a set of states $\left|\varphi_{i}\rangle \right . $ 
(where the index i runs over the 
degrees of freedom of the system) that form a basis and the most general state of such a 
system can be written as \ket{\Psi}$=\sum_{i}c_{i}$\ket{\varphi_{i}}. The system is said 
to be in a state \ket{\Psi} which is 
a linear superposition of the basis states $\left|\varphi_{i}\rangle \right . $  with weighting 
coefficients $c_{i}$ that can in 
general be complex. At the microscopic or quantum level, the state of the system is
described by the wave function \ket{\Psi}, which in general appears as a linear superposition of 
all basis states. This can be interpreted as the system being in all these states at once. It is 
known that the tubulin dimer undergoes conformational changes as a result of a shift in 
the localization of the electron orbitals in its hydrophobic pocket. Therefore, a superposed 
state of the tubulin dimer would have the interpretation of the dimer {\it being in both of its 
allowable conformational states at the same time} something which is not allowable 
classically. 

The coefficients $c_{i}$ are called the probability amplitudes and $\left|c_{i}\right|^{2}$ gives the 
probability that  $\left|\Psi\rangle \right .  $ will collapse into state 
$\left|\varphi\rangle \right . $ when it decoheres 
(interacts with the environment). By simple normalization we have the constraint that 
$\sum_{i}\left|c_{i}\right|^{2}=1$. 
This emphasizes the fact that the wavefunction describes a {\it real, physical system}, which must 
be in one of its allowable classical states and therefore by summing over all the 
possibilities, weighted by their corresponding probabilities, one must obtain unity. 
Further, this fact stresses that quantum mechanics is not simply an alternative treatment of 
such two-state systems as tubulin dimers but rather it is the {\it correct} mathematical 
treatment. A quantum mechanical treatment of the tubulin dimer does not rely on 
approximations contrary to the case when a tubulin dimer is treated using classical 
electrodynamics, thermodynamics and statistical physics (which is the usual approach in 
biophysical investigations of protein molecules).

Note that at the macroscopic or classical level, a quantum two-state system can 
only be in a single basis state. For instance, the quantum position (energy) of an electron 
can be in a superposition of two different orbitals (energies) while in the classical case 
this is impossible. Equally, the tubulin dimer can only be experimentally observed 
(measured) in one of its two allowable conformations. 

\subsection{Role of Coherence \& Entanglement in Recall \& Binding}
A quantum system is {\it coherent} if it is in a linear superposition of its basis states. If 
a measurement is performed on the system and this means that the system must somehow 
interact with its environment, the superposition is destroyed and the system is observed to 
be in only one basis state, as required classically. This process is called {\it reduction} or 
{\it collapse} of the wavefunction or simply decoherence and is governed by the form of the 
wavefunction \ket{\Psi}.

In this notation, the probability that a quantum state \ket{\Psi} will collapse into a basis 
state \ket{\varphi_{i}} is written in terms of the inner or scalar product 
$\left|\right . $\bra{\varphi_{i}} $\Psi\rangle  \left|^{2} \right .$ which is analogous 
to the familiar dot product between two vectors $(\vec{a}\cdot\vec{b})$. The simplest system which 
would be analyzed as described above, would be a two-state system. For instance, an 
electron (spin--$\frac{1}{2}$) system, where we are interested in measuring the electron's spin in a 
specified direction (customarily the z-axis). The general notions of the simplified 
mathematical treatment that follows can also be applied to the tubulin dimer\protect\cite{nan1}. The actual 
experimental setup for measuring the orientation of the spin of an electron is called a 
Stern-Gerlach (SG) apparatus described in detail in a number of sources, e.g. \protect\cite{sak1}. When inserted 
into an SG magnet, the electron can either register as spin-up (\ket{1}) or spin-down (\ket{0}). In 
this system, the wavefunction is a distribution over the two possible values and a coherent 
state \ket{\Psi} is a linear superposition of \ket{1} and \ket{0}. One such state can be:
\hspace{0.3cm}
\ket{\Psi}=\(\frac{2}{\sqrt{5}}\)\ket{1}+\(\frac{1}{\sqrt{5}}\)\ket{0}
 
As long as the system remains in a coherent state, we cannot say that the electron 
is in either the up- or down-spin states. In a counter-intuitive sense, it is in both states at 
once. Classically, the electron can only be in one state, so if measured, the system 
decoheres to give spin up with probability: 

\centerline{
$\Big|$\bra{1}$\Psi\rangle \Big|^{2}=\left(\frac{2}{\sqrt{5}}\right)^{2}=80\%$
}

and spin down with probability 20\%.

This simple two-state quantum system is used as the basic unit of quantum 
computation and is referred to as a quantum bit or qubit.

In classical, digital computers, the basic unit of computation is a bit. A bit can 
have the value 1 or 0. Digital computers encode information using arrays of bits in the 
form of billions of solid-state transistors integrated to form microchips. The voltage at 
each transistor can take two values making it a 1 or a 0. Computation i.e. manipulation of 
information, is performed by logic gates which follow Boolean algebra rules. In quantum 
computers, the logic gates are replaced by quantum operators. Operators on a Hilbert 
space describe how one wavefunction is changed into another. An eigenvalue equation 
shows the action of operators. For instance, an operator A acting on one of its own basis 
states \ket{\varphi_{i}} will produce the same state multiplied by its eigenvalue $a_{i}$,
namely: $A$\ket{\varphi_{i}}$=a_{i}$\ket{\varphi_{i}}

The solutions of the eigenvalue equation \ket{\varphi_{i}} are called the eigenstates and can 
be used to construct a basis. In quantum mechanics, we assign operators to all the 
physical properties of a system (such as position, momentum, energy) and the eigenvalues 
of these operators give us the allowed physical values of those properties.

As it will shortly become important, note that while {\it interference} is a commonly observed 
classical wave phenomenon, it has 
also been experimentally shown to apply to the probability waves of quantum mechanics. 
For a simple theoretical example, consider the initial wavefunction for the spin--$\frac{1}{2}$
electron system described earlier.  Using the conventional vector assignment,

\ket{1}=\MH{1}{0},\hspace{1cm}\ket{0}=\MH{0}{1}

we can rewrite our wavefunction in vector form as:

\ket{\Psi}=\(\frac{1}{\sqrt{5}}\) \MH{2}{1}

When acted upon by some operator A, where A is defined to be:

$A=\frac{1}{\sqrt{2}}\ \left(\begin{array}{cc}1&1\\1&-1\end{array}\right)$

the result is:

%\centerline{
%A\ket{\Psi}=\(\frac{1}{\sqrt{2}}\) \left(\begin{array}{cc}1&1\\1&-1\end{array}\right)
%\frac{1}{\sqrt{5}}\ \MH{2}{1}=\(\frac{1}{\sqrt{10}}\) \MH{3}{1}=
%\(\frac{3}{\sqrt{10}}\) \ket{1}+\(\frac{1}{\sqrt{10}}\) \ket{0}
%}

\beqn
A | \Psi \rangle = \frac{1}{\sqrt{2}}
\left( \begin{array}{cc} 1 & 1 \\ 1 & -1 \end{array} \right) 
\frac{1}{\sqrt{5}}\left( \begin{array}{c} 2 \\ 1 \end{array}  \right) 
=\frac{1}{\sqrt{10}}\left( \begin{array}{c} 3 \\ 1 \end{array}  \right) 
= \frac{3}{\sqrt{10}} | 1 \rangle + \frac{1}{\sqrt{10}}| 0 \rangle \, \nolabel
\eeqn

%A\kket{\Psi}=\(\frac{1}{\sqrt{2}}\) \left(\begin{array}{cc}1&1\\1&-1\end{array}\right)
%\frac{1}{\sqrt{5}}\ \MH{2}{1}=\(\frac{1}{\sqrt{10}}\) \MH{3}{1}=
%\(\frac{3}{\sqrt{10}}\) \kket{1}+\(\frac{1}{\sqrt{10}}\) \kket{0}
%\nolabel
%\eeqn
 
Note that as a result of the action of A on our initial state \ket{\Psi}, the amplitudes of 
the spin-up and spin-down states have changed. The operator has made the wavefunction 
interfere with itself and its constituent parts experienced the analogue of classical 
interference so that the up state interfered constructively while the down state destructively.

{\it Entanglement} on the other hand, is a purely quantum phenomenon and has no 
classical analogue. It accounts for the ability of quantum systems to exhibit correlations 
in counterintuitive "action-at-a-distance" ways. Entanglement is what makes all the 
difference in the operation of quantum computers versus classical ones. We will present a 
short mathematical description here without using density matrix formalism.

If we wish to describe the state of two  electrons (spin--$\frac{1}{2}$), or equally, the state of two 
tubulin molecules, we may use Dirac bra-ket notation where the first entry in a ket refers 
to the state of the first electron (conformation of first dimer) while the second entry refers 
to the second electron (second dimer). For instance, let us take a quantum state \ket{X} made 
up of two electrons where the first is in the spin-down state with certainty while the 
second is in a coherent state of spin-up and spin-down with equal probability. 

\centerline{
\ket{X}=\(\frac{1}{\sqrt{2}}\)\ket{00}+\(\frac{1}{\sqrt{2}}\)\ket{01}}

another may be state \ket{\Psi}:

\centerline{
\ket{\Psi}=\(\frac{1}{\sqrt{2}}\)\ket{00}+\(\frac{1}{\sqrt{2}}\)\ket{11}}

and a state \ket{\zeta} could be:

\centerline{
\ket{\zeta}=\(\frac{1}{\sqrt{3}}\)\ket{00}+\(\frac{1}{\sqrt{3}}\)\ket{01}+\(\frac{1}{\sqrt{3}}\)\ket{11}}
 
where all states are indexed by the state labels 00,01,10,11.

These three states are different from each other in the sense that although \ket{X} can 
be factorized using normal tensor product ($\bigotimes$) as follows:

\beqn
|X \rangle = \frac{1}{\sqrt{2}}| 0 \rangle  \bigotimes \left( \, | 0 \rangle + | 1 \rangle \, \right) 
\nolabel
\eeqn

\ket{\Psi} cannot be factorized. States that {\it cannot be factorized are said to be entangled 
states}. Note that \ket{\zeta} can be factorized in two different ways but not completely. There are 
degrees of entanglement as states can be less or more entangled, depending on whether 
they are completely, partially or not at all factorizable and the three states \ket{\Psi}, \ket{\zeta} and 
\ket{X} demonstrate this.

Entanglement gives "special powers" to quantum computers because it gives 
quantum states the potential to exhibit and maintain correlations that cannot be 
accounted for classically. Correlations between bits are what make information encoding 
possible in classical computers. For instance, we can require two bits to have the same 
value thus encoding a relationship. If we are to subsequently change the encoded 
information, we must change the correlated bits in tandem by explicitly accessing each 
bit. {\it Since quantum bits exist as superpositions, correlations between them also exist in 
superposition}. When the superposition is destroyed (e.g. one qubit is measured), {\it the 
correct correlations are instanteaneously ``communicated'' between the qubits} and 
this communication allows many qubits to be accessed at once, 
preserving their correlations, something that is absolutely impossible classically. 

"Software" that makes use of this possibility has already been 
developed in the form of factorization\protect\cite{shor1}, sorting\protect\cite{gro1} and
learning and memory\protect\cite{ven1}  algorithms. 
This communication of correlations is a manifestation of the well-known Einstein--
Podolski--Rosen (EPR) paradox. A simplified example follows.

Consider the case of the pion ($\pi^0$), a neutral elementary particle of spin 0 (i.e. 
internal angular momentum 0). Pions decay spontaneously into two oppositely polarized 
photons. Photons carry angular momentum in their helicity. Since ($\pi^0$) decay is spontaneous 
i.e. no external forces have acted on the system, angular momentum conservation requires 
the decay products to have the same total angular momentum as the decaying particle --in 
our case a sum of zero. This means that if one photon is detected with helicity +1 the 
other must have helicity -1 to conserve angular momentum. We can write the entangled 
state of the two emerging photons as:

\centerline{
\ket{\gamma_1,\gamma_2}=\(\frac{1}{\sqrt{2}}\)\ket{-1,1}+\(\frac{1}{\sqrt{2}}\)\ket{1,-1}
}
 
where the subscripts 1 and 2 refer to the first and second photons respectively and 
the +/- 1 entries in the kets refer to each photon's helicity. This state is completely 
entangled as it cannot be factorized to give separate states for each photon. This indicates 
that if the product photons are isolated from the environment and separated 
macroscopically (say by letting them move apart inside an optical fiber), measuring one 
photon's polarization will immediately determine the polarization of the other by 
collapsing the entangled wavefunction instantaneously and non-locally. 

\newpage
\section{EXPERIMENTS}

\subsection{Experiments}
Our novel phenomenological approach to understanding the role of MTs in 
information processing has produced several theoretical predictions, which we aim to 
support with experimental evidence. Ideally, we would like to test our predictions {\it in 
vivo} by examining the effects that learning and memory encoding have on the MTs of a 
living animal. One way to do this would be to disrupt MTs by using mutations. 
However, it is impossible to use animals that harbor mutations that 
change functional aspects of MTs as the MTs' correct function is essential 
for the viability of an organism. Instead, we have designed and are performing several 
indirect neurobiological experiments. We anticipate to obtain the first-ever experimental 
results designed to test the quantum properties of living matter. 

\subsection{Description of the experimental system}
We utilize a well-established, olfactory conditioning protocol to teach 
Drosophila melanogaster fruit flies to avoid certain odors contingent upon negative 
reinforcement by electric shock.  Following behavioral conditioning and ascertaining 
acquisition of information, the flies are fixed, their heads sectioned and stained 
(immunochemically) for the distribution of tau, or microtubule-associated-protein-2 
(MAP-2) in the mushroom bodies (an area of the fly's brain essential for information 
correlation and memory formation). We are interested in determining whether, as 
predicted by our model, the distribution of tau and/or MAP-2 will change as a result of 
memory encoding.

\subsection{Why Use Drosophila?}
The Drosophila melanogaster fruit fly has long being favored by experimental 
biologists for numerous reasons including its relatively simple genetic makeup 
(genome) and quick generation time, powerful classical and molecular genetics and 
their ability to learn and remember a variety of tasks. However, Drosophila is simply 
the ideal system for our research for a different reason.  In order to track redistribution 
of tau and MAP-2 in the neurons we must be able to differentiate between the various 
parts of the neuron such as the dendrites, axons, axonal projections and somata. In 
humans and other mammals, the neuronal organization is such, that multiple neurons 
and neuronal types are involved in a given process forming an extensive complex 
network of axons and dendrites. As a result, it is very difficult to locate specific parts of 
individual neurons and stain selectively to track changes in distribution of a particular 
protein.  In Drosophila on the other hand, mushroom body neurons represent a highly 
ordered, tightly packed bundle (see figures 5 and 8). 

The mushroom bodies are bilateral clusters of about 2500 neurons, situated in 
the dorsal and posterior cortex of the Drosophila brain (figure 5).  The dendrites of all 
mushroom body neurons aggregate to form a distinctive structure just ventral to the cell 
bodies where inputs arrive conveying sensory information.  The axons of these neurons 
bundle together (fasciculate) and project to the anterior of the brain.  There, they 
bifurcate, with some axonal processes extending medially and others projecting 
dorsally\protect\cite{dav1}  (figure 5).  Flies that lack mushroom bodies are able to smell, but totally 
unable to learn the olfactory associative learning task\protect\cite{dav2,skou1}.

\begin{figure}
\epsfxsize=5in
\centerline{\epsffile{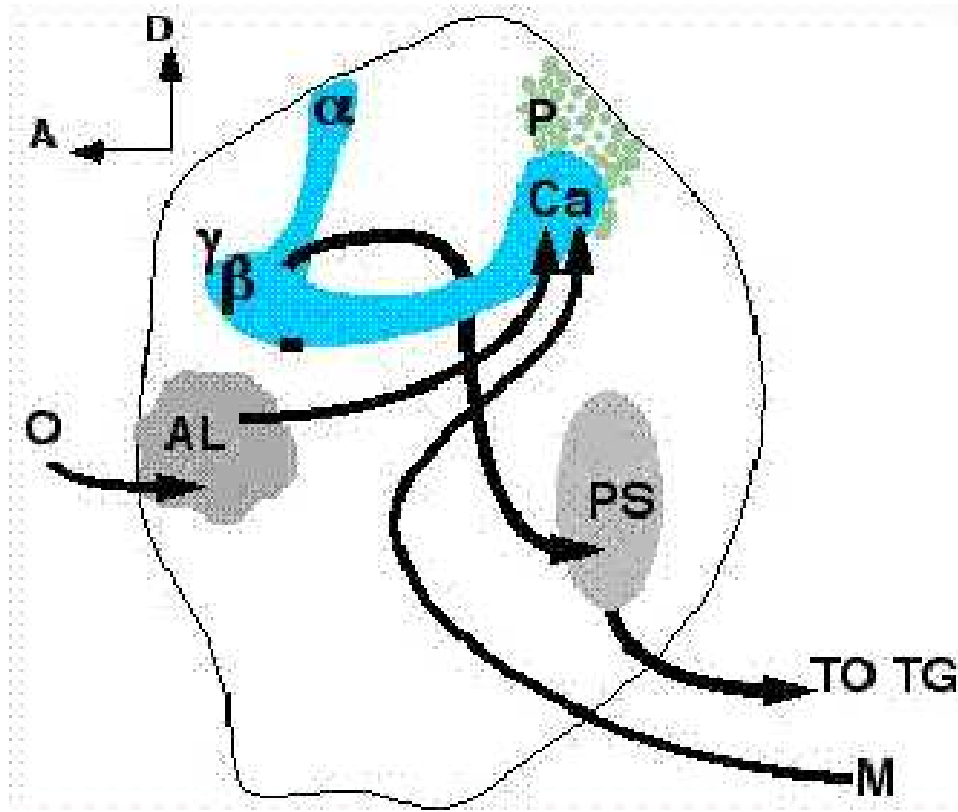}}
\caption{A schematic diagram of a saggital section through the fly brain. A: 
anterior and  D: dorsal. AL: 
Antennal lobe. PS: Posterior Slope. P: mushroom body perikarya. Ca: mushroom
body calyx. The lobes 
are labeled $\alpha$, $\beta$ and $\gamma$. The hypothetical flow of information is 
represented by the arrows. O: Olfactory 
information. M: mechanosensory information from thoracic ganglia. TO TG: information output from the 
posterior slope to the thoracic ganglia.}
\end{figure}

Therefore, the Drosophila system enables us to target a particular set of neurons 
easily identifiable and well described in their properties. This is essential for our 
analysis of bulk movement of microtubule-associated proteins within neurons.

\subsection{Conditioning Protocol}
A qualitative description of our protocol\protect\cite{dav1,dav2,skou1,tul1} for training wild type 
Drosophila Melanogaster fruit flies follows. Drosophilae are naturally attracted or 
repulsed by different odors with a variety of affinities. We use the standard, negatively 
reinforced associative learning paradigm which couples olfactory cues with electric 
shock to condition flies. We used two equally aversive odorants: 3-Octanol (OCT) and 
Benzaldehyde (BNZ). The training apparatus consists of a training chamber and a 
selection maze. The maze is normalized by adjusting the concentration of odorant. Once 
normalized, wild type, naive (i.e. untrained) flies choose to enter one of two identical 
tubes smelling of OCT and BNZ respectively, with a probability of 50\% since they 
avoid both odors equally. 

Training, or rather conditioning, of the flies takes place as follows. A batch of 
wild type, naive flies (numbering between 50 and 60) are collected under light 
anesthesia (using $CO_{2}$) and 12-24 hours later are left in the dark for one to two hours. 
The entire procedure of conditioning the flies takes place in a temperature- and 
humidity-controlled darkroom. This is done in order to isolate the effects of olfactory 
stimulation from visual stimulation. Once the flies have been acclimated to the 
darkroom, they are inserted into training chamber A whose walls are electrified by a 
signal generator set to 92.0V.  The flies receive twelve electrical shocks (of duration 
1.25sec each) for one minute. During this time, the chamber is filled with air containing 
OCT. The flies are given 30 seconds to rest while the air is being cleared of odorants 
and are then given the opposite (control) odorant (in this case BNZ) for another minute 
in the absence of electrical shocks.  A rest period of 30 seconds follows after which the 
flies are tested for acquisition of information. They are inserted into the selection maze 
and given the choice of entering a chamber smelling of OCT or an identical one 
smelling of BNZ. For control and consistency purposes, the experiment is done 
simultaneously in apparatus B with the shock-associated and control smells reversed 
while everything else remains identical. We define a "trained" fly as one that has 
chosen to go into the chamber filled with the control odor after given the choice for 90 
seconds. The procedure is illustrated in figure 6 below.

\begin{figure}
\epsfxsize=4in
\centerline{\epsffile{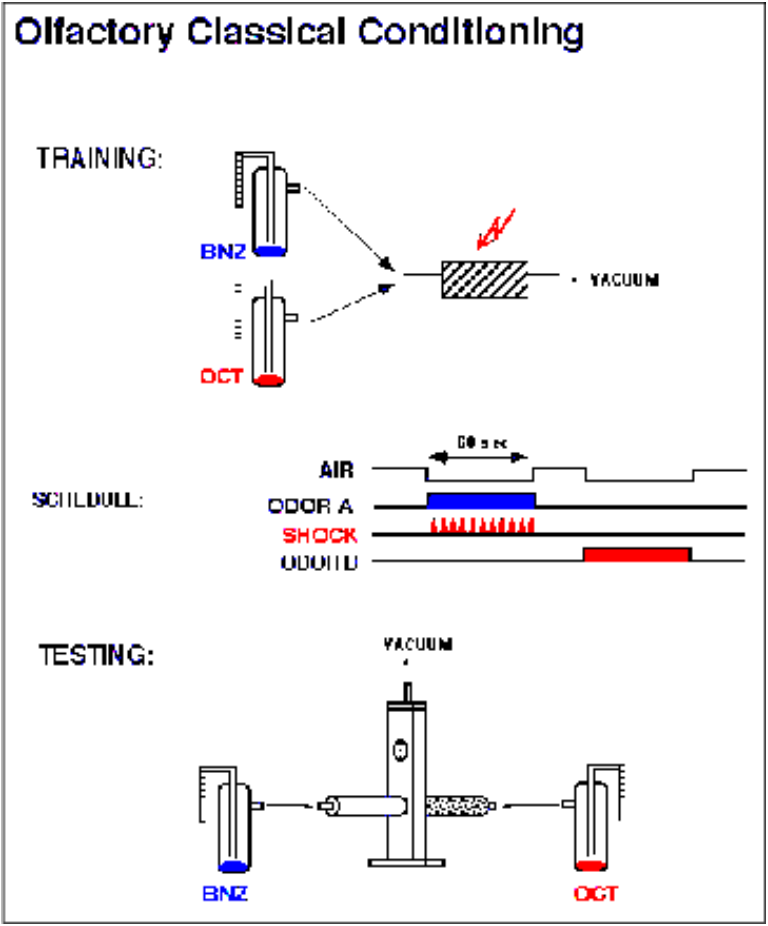}}
\caption{A schematic diagram of the training and testing apparati and schedule for the negatively 
reinforced olfactory conditioning protocol}
\end{figure}

It is observed that following training, a good percentage of the flies choose to 
avoid the smell that was present when they received the electrical shocks. The 
percentage is calculated as a normalized performance index (PI) where

$PI=\left(\frac{trained-untrained}{total}\right)\times 100$.

Typical PI values for our experiments have been between 75 and 
90 giving us confidence that the flies have truly learned to associate the stimuli. This 
procedure alters the probability of response of the flies to the stimuli.  Re-testing the 
flies that made the correct choice producing a PI of 90 will not result in 100\% of the 
flies avoiding the shock associated odor, but rather will result in a distribution 
producing a PI of 90 again.  Therefore, the behavioral changes of the flies parallel the 
probabilistic response of neuronal firing.

The procedure we followed is typical of associative learning "Pavlovian" 
conditioning for behavioral experiments involving a variety of animals and more details 
can be found in the literature\protect\cite{woo1,tul1}.  

\subsection{Fixation, Sectioning and Staining}
Once the flies have made their choice, those that made the 'correct' choice are 
immediately killed by submersion into a fixative solution without subjecting them to 
anesthesia.  An equal number of naive flies that have not been exposed to the training 
apparatus but are otherwise identical to the trained ones are also fixed.  We used three 
different fixing protocols of different fixation strength each\protect\cite{tul1}. Following fixation the 
flies were dehydrated through a series of ethanol baths (0-100\%) and Methylbenzoate 
preparing them for embedding in Paraffin, decapitating and sectioning. It is clear that 
this experimental approach is likely to capture differences in the distribution of tau and 
MAP-2 between trained and naive flies and consequently provides us with a way of 
testing whether their distribution is affected by training as predicted by the quantum 
brain hypothesis and also suggested by recent studies in rodents\protect\cite{woo1,woo2,woo3}. There are however, 
a number of complications.  In order for the distribution of microtubule associated 
proteins to be seen in the microscope, one must bring it up from the background by 
immunochemical methods which use in situ chemical staining reactions to indicate the 
localization of the proteins within the cell.  This involves obtaining antibodies (which are 
also proteins) that selectively and specifically attach to tau or MAP-2. Following standard 
immunochemical procedures, the antibodies become linked to chromogenic (staining) 
substances that allow visualization of the distribution of the protein under investigation 
in the tissue of interest. 

Note that there is an important underlying assumption here: the fixative solution 
is required to fix the tissue as it was at the instant of death so that all proteins in the 
neurons are permanently bound to their last location before death. It further assumes 
that embedding in Paraffin will not affect the binding of MAPs to MTs or the structure 
of MAPs themselves. Such changes may alter the structure of MAPs in ways that will 
make them no longer identifiable by the antibodies.  Historically, fixation complications 
have been circumvented by using a variety of fixatives and fixation conditions with 
good results despite the lack of complete theoretical understanding as to the exact action 
of the fixatives. 

Furthermore, due to resolution and staining limitations, it is not possible to 
directly test the GSM (guitar string model) unless there is sufficient relocation of 
MAPs.  At least in this {\it modus operandi}, we have had no choice but to assume that the 
training has been sufficiently intensive so that the result of encoding this memory was 
to dramatically change the MAP-2 distribution in a large number of neurons.

\subsection{Results}
We have been successful in our initial experiments to train flies and test their 
learning and memory for up to 6hrs.  This is an essential point as we are not certain of 
the exact timing of the proposed redistribution of microtubule associated proteins.  
However, our initial attempts to localize MAP-2 within the fly brain have been 
unsuccessful. It must be noted that in our experiments, we used monoclonal antibodies 
which only recognize one binding site at the target protein and if that site is  "buried" in 
the secondary structure of the protein, the antibody will not bind. A solution to this 
problem is to use polyclonal antibodies, which have a number of binding sites on their 
target protein.  We are currently in the process of trying a number of anti-MAP-2 
polyclonal antibodies to select the one that best reveals the MAP-2 distribution in the 
fly brain.

A more interesting interpretation of these initial results is that the antibody 
binding sites on MAP-2 are "masked" due to its interaction with the MTs, but upon 
training and during the proposed re-distribution, these sites may become available for 
detection. We are in the process of addressing this possibility by training flies and 
fixing them at different times (0 to 360 minutes) past training, to investigate whether 
these proposed sites become available which would be evidenced by immunological 
staining.

\subsection{Experiment $\#$2 Basics}
Tau is another microtubule associated protein. In humans, it plays a role similar 
to MAP-2 and it seems that tau is of paramount importance in keeping axonal MTs 
parallel and aligned.

We have obtained transgenic flies that will express human tau-protein in their 
mushroom bodies. This is important as tau has long being implicated in the encoding of 
human memory and it has recently been shown that mutations in the human NC-17 tau 
gene are one of the causes of Alzheimer's Disease (AD)\protect\cite{vog1}. In fact, earlier 
theoretical research by this group has led us to assert this prior to it becoming well 
accepted. We had claimed that subneural abnormalities such as Neurofibrillary Tangles 
(NFTs) and abnormally phosphorylated tau are the main causes of AD symptoms, rather 
then the other way around. NFTs are axonal MTs that have lost their structural integrity 
due to the inability of mutated or hyperphosphorylated tau to hold them in parallel and 
as a result have been tangled up and are unable to function properly.

\subsection{Motivation, Relevance to Alzheimer's Disease}
By inducing flies to express the human tau-gene specifically in their mushroom 
bodies, we anticipate that we will in fact be replacing, at least to some extent, the MAP 
the fly actually uses to hold its MTs together by human tau-protein. We do not know a 
priori what to expect, as the flies can exhibit an increase or decrease in learning 
performance, or there might be no overt phenotype.  In the case that there is observable 
phenotype in their learning, we will be able to deduce something about the role that 
MAPs in general play in memory encoding. We already know that the introduced tau 
gene is not lethal and a preliminary investigation does not indicate anomalies in general 
feeding, mating, or circadian behaviors of the flies. Furthermore, assuming that tau will 
play a similar role as it does in humans, we will be able to test whether overproduction 
of tau in older flies makes them susceptible to "dementia" in the form of a 
neurodegenerative disease such as Alzheimer's. In the case that the flies exhibit 
learning deficits, we will examine their brains for histological hallmarks such as NFTs. 
Alternatively, the introduction of human tau may reduce or eliminate the observed age 
dependent decline in the learning capability of fruit flies.

\subsection{Protocol}
How does one go about persuading a fly to create a human protein in its brain?  
This process is called {\it directed gene expression} and uses genetic engineering to force the 
expression of genes in specific tissues, even if they are alien to the organism.  This 
method also allows turning gene expression on or off at specific times. The main idea is 
to have two genetically manipulated lines the first of which contains a gene of choice 
(human tau in our case) fused to and under the direction of an upstream activating 
sequence (UAS) activated only by the presence of its unique, selective and specific 
activator protein GAL4 in the same cell. To generate lines expressing GAL4, the GAL4 
gene is inserted randomly into the fly's genome in front of various genes expressed in 
specific tissues at specific developmental times (temporal control) due to the action of 
their native enhancers dictating this expression pattern.  The GAL4 transgene "usurps" 
these native enhancers, resulting in its tissue and temporal specific expression.  A 
GAL4 target gene (UAS-tau) will remain silent in the absence of GAL4.  To activate 
the target gene, the flies carrying the UAS-tau are crossed to flies expressing GAL4 and 
thus in their progeny, the UAS-tau transgene will be expressed in the tissue and 
temporal specific pattern specified by the GAL4. This is illustrated in figure 7 below.
 
\begin{figure}
\epsfxsize=5in
\centerline{\epsffile{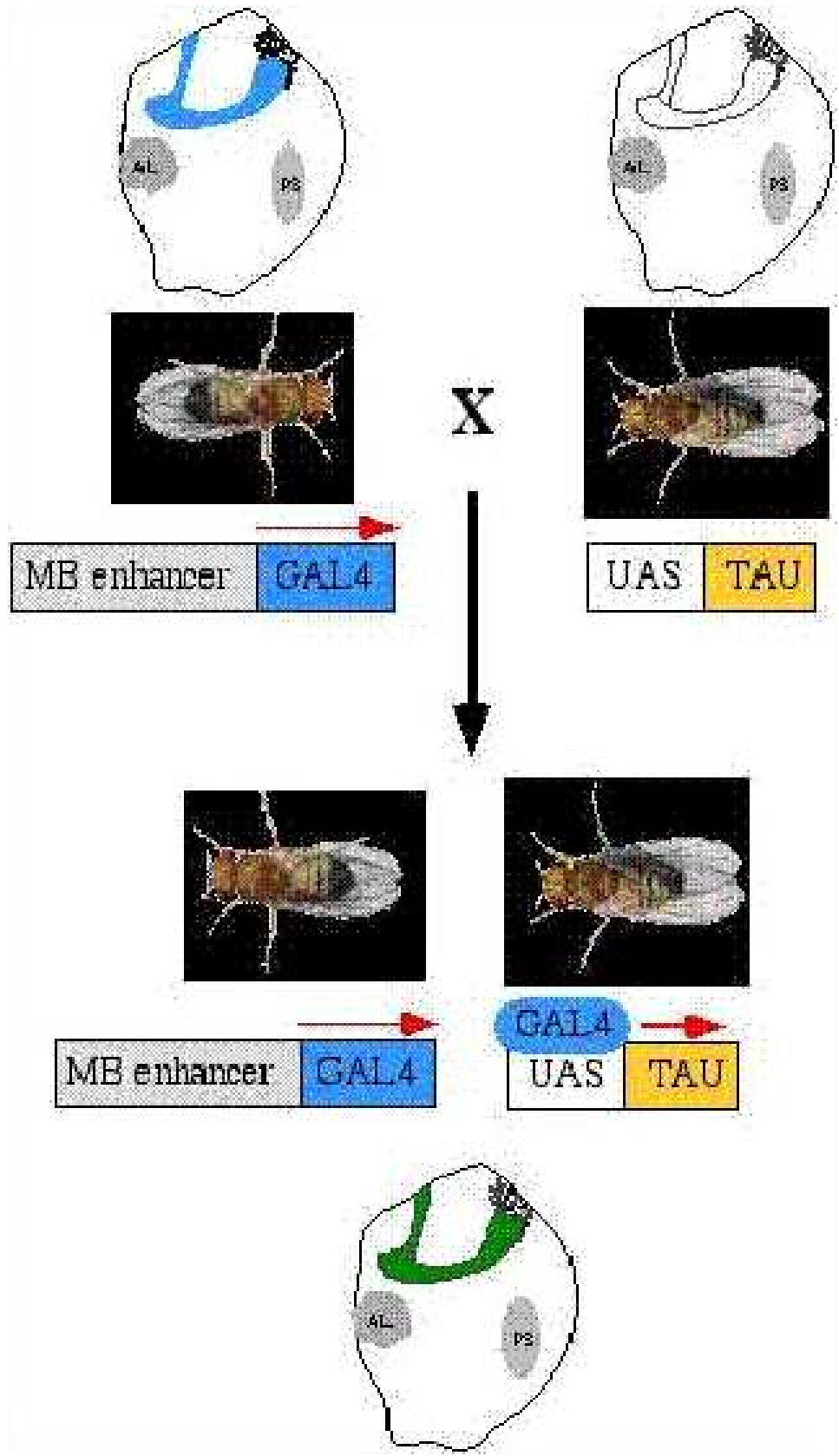}}
\caption{Schematic illustration of genetic crosses to generate flies expressing human 
tau in their mushroom bodies.}
\end{figure} 

\subsection{Fixation, Sectioning and Results}
This experiment is currently underway. To ascertain that we have flies 
expressing human tau in their mushroom bodies we must first test whether the GAL4 
"driver" line directs expression in the mushroom bodies as advertised. To do this, we 
cross flies that contain the GAL4 gene with flies that contain another gene whose 
activity can be readily monitored (reporter gene) by histological methods.  We have 
used the bacterial beta-galactosidase gene (UAS-LACZ). 

Flies that are the progeny of GAL4xLACZ will have beta-galactosidase activity 
in their mushroom bodies visualized as blue pigment. This provides us with a simple 
test of where the actual tau-gene is going to be expressed once activated in a GAL4xtau 
cross. 

The sectioning procedure employed here is different from the one for 
experiment $\#$1.  The naive fly heads are cryo-sectioned by freezing to $-20^{o}C$ and an 
embedding gel is used instead of chemical fixation which would destroy the activity of 
the reporter gene.  The staining is provided by the activity of the reporter LACZ gene 
which converts a colorless substrate into a blue precipitate within the tissue where the 
reporter is expressed. The results of this preliminary experiment are encouraging as it is 
seen beyond doubt that the mushroom bodies as well as certain other sections of the fly 
brain do indeed express LACZ indicated by the blue color in the sections. This is 
illustrated in figure 8.

\begin{figure}
\epsfxsize=5in
\centerline{\epsffile{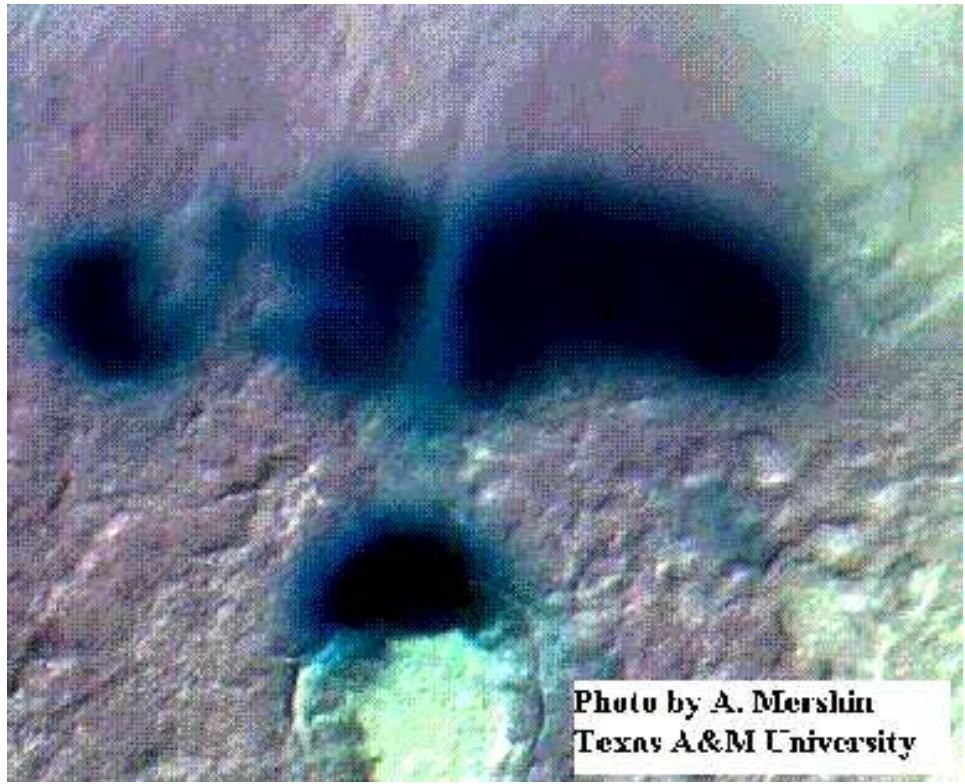}}
\caption{Frontal cryo section of fly brain expressing LACZ. Dark (blue) staining indicates directed 
expression of the LACZ gene. The stained structure is histologically identified as a mushroom body.}
\end{figure} 
 
The next steps of this experiment are to first investigate the learning phenotype 
of the GAL4xtau flies and then proceed with the sectioning and staining methods 
described for experiment $\#$1 using anti-tau antibodies to visualize potential changes in 
the distribution of tau immediately after training and at later times to assess changes due 
to memory formation. Figure 9 shows our preliminary immunohistochemical results 
where the dark staining corresponds to expression of tau. The flies used for this were 
naive, i.e. have not been exposed to the training apparatus. 

\begin{figure}
\epsfxsize=5in
\centerline{\epsffile{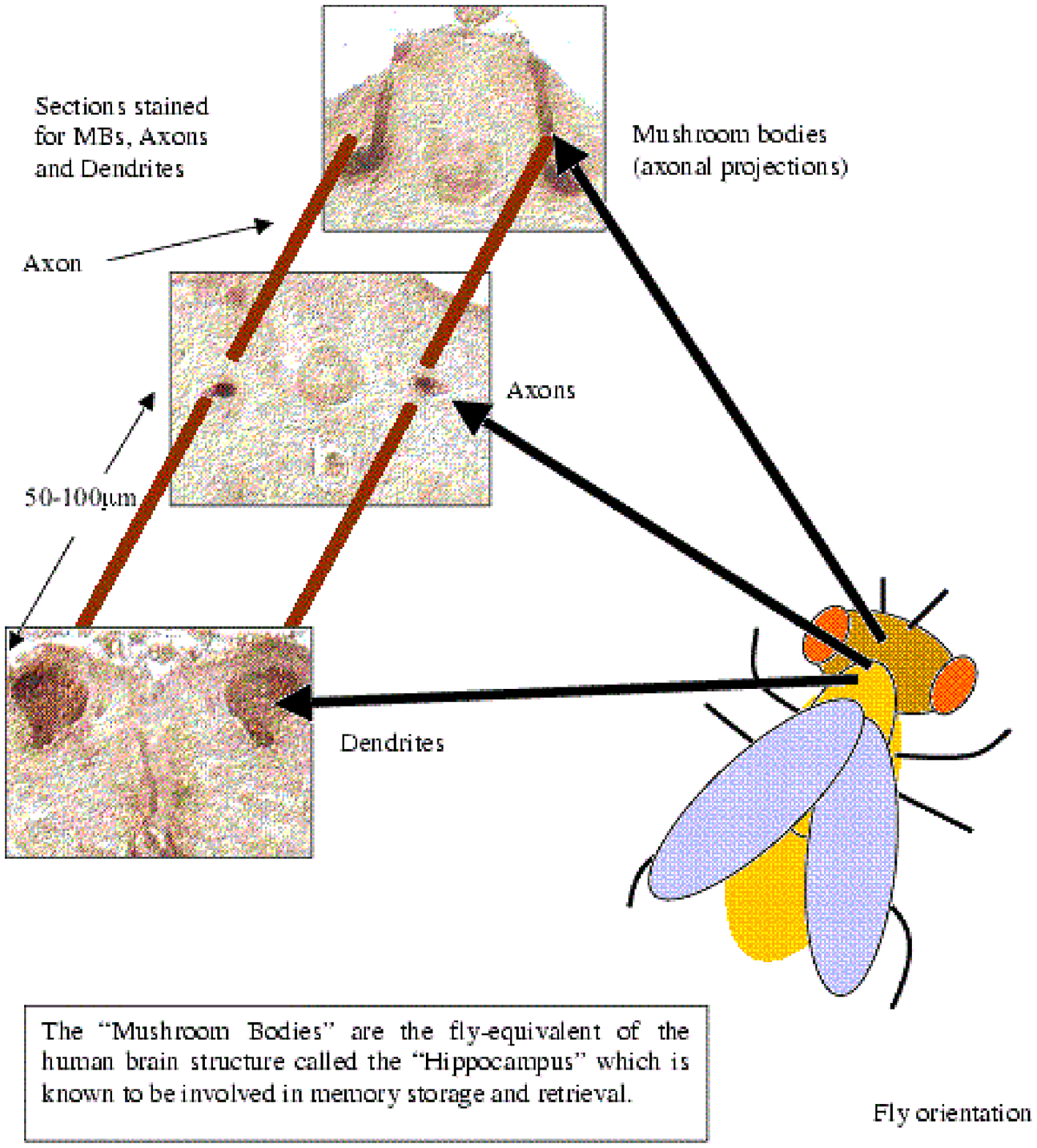}}
\caption{Results of trangenic,tau-expressing, fly-brain sections and tau-staining}
\end{figure}

\newpage
\section{SUMMARY, DISCUSSION AND CONCLUSION}
\subsection{Summary}
This review gives an overview of the new field of 'quantum physics motivated 
neurobiology'. A biological model of memory was presented that stressed the role of the 
cytoskeleton in encoding memory. The quantum brain hypothesis was outlined and 
some phenomenological aspects discussed. Some fundamental aspects of cell and 
molecular neurobiology were presented, concentrating on microtubules and 
microtubule-associated-proteins. The reasons why we believe quantum coherence is a 
realistic possibility even at the scale of MTs and even at temperatures of order $37^{o}C$ 
were outlined in Section 3. Quantum computation was discussed and a brief example of 
entangled states of qubits was presented in Section 4. In Section 5, two experiments 
targeted at indirectly testing the phenomenological predictions derived from our 
quantum theory of brain were described and some preliminary results presented. 

\subsection{Discussion}
It cannot be stressed enough that this line of research is suffering from a total 
lack of {\it in situ}, direct quantum mechanical experiments. Recently published\protect\cite{sac1} and some 
yet unpublished efforts by Zioutas et. al. concentrate on showing that MTs are indeed 
ferroelectric. They use detection of hypothetical ferro-to-paraelectric state phase 
transitions due to temperature changes to investigate the conjectured ferroelectric 
properties of MTs. They employ a novel approach where they try to detect 
electromagnetic radiation (of order some GHz) emitted by MTs as a result of the 
expected phase transition. 

The theoretical aspects of this work have been criticized by some physicists 
claiming that quantum coherence is impossible in the $37^{o}$C biological 
environment\protect\cite{teg1}. It 
has been argued that if one sets up a wavefunction describing a whole neuron including 
its membrane and the surrounding ions it will decohere with decoherence times no 
longer than $10^{-19}$ s due to collisions between ions and $H_{2}O$ molecules as well as long 
range Coulomb interactions. This certainly seems plausible and biological manifestation 
of quantum mechanical effects will not be observed since the biological dynamical 
timescale is of order $10^{-4}-1s$. This argument is flawed in assuming that the entire neuron 
must be in a coherent state of "firing-and-not-firing" and thus for a typical axon up to 
order $10^{6}$ Na$^+$ ions must be in a superposition of being in-and-out of the axon and in 
coherence with each other and the membrane. The firing or not of the neuron is an 
emergent property and it seems unlikely that quantum mechanics will apply directly to 
the generation and propagation of APs let alone the bulk dynamics of extracellular ions. 
{\it Our model does not suggest superposition of ions and membrane}. 

Further, it has been argued\protect\cite{teg1} that even MTs will have difficulty sustaining 
coherence and an estimate of the decoherence time is given to be of order $10^{-13}sec$. The 
dominant decoherence mechanism for these considerations is assumed to be Coulomb 
interactions with closest neighboring ions. Screening effects are not considered by 
arguing that the closest ions will be the ones doing the screening and they are also the 
ones effecting decoherence. This, it is claimed, does not allow information propagation 
by the mechanisms suggested. However, as described in Section 3, the particular 
ordered arrangement of $H_{2}0$ molecules inside and outside the MT as well as the 
presence of the K-code may well protect MTs from such decohering mechanisms\protect\cite{mn1}. 
It is an ongoing effort\protect\cite{sat1} to further 
unveil how these properties will affect decoherence-time 
estimates. Decoherence may well be effected by strong electromagnetic interactions 
resulting from neurotransmitter binding and this is exactly the kind of decoherence 
mechanism which should be most favored by biologists since neurotransmitter action 
and consequent AP propagation along the membrane must somehow "reset" the system 
to where it can receive the next "information package". It remains to be seen if by 
further theoretical analysis and biophysical experimentation, the decoherence time can 
be conclusively brought up to match the dynamical time scale. 

A natural extension of the quantum hypothesis is that there is place for quantum 
coherent effects in other, non-neural cells. In fact anything with a cytoskeleton-like 
structure, any protein whose function depends on electron mobility (and this includes 
all known proteins) can be treated as a fundamentally quantum mechanical object. 
Whether there are observable emergent properties depends on the system at hand but it 
seems that the difference between neural and ordinary cells is made by the 
characteristically ordered and long MTs in the axons and dendrites. It remains to be 
seen what role quantum mechanics is going to play in the molecular biology of the 
future. 

\subsection{Conclusion}
This review is an account of the initial steps of what we expect to be a long 
process of theoretical and experimental research in this field. What we have achieved so 
far is to design and conduct the first ever experiments capable of indirectly testing some 
of the predictions of the quantum brain hypothesis. It cannot be stressed enough that the 
experiments described here can only tell us whether the quantum hypothesis, at least in 
its present formulation, is {\it wrong}. Even if all of the theoretical predictions are shown to 
hold in the laboratory, these results can have other, more conventional, interpretations. 
For instance, if we are able to show a definite redistribution of MAPs as a result of 
learning, it can be argued that this changes neural cells in ways which do not directly 
depend on quantum coherence e.g. axonal transport can be altered thus affecting 
synaptic weight and effectively training the neuron.  As discussed earlier, in order to 
discover at which level the transition from quantum to classical takes place, direct 
quantum mechanical experiments on the MT system are needed but those seem to be 
quite far ahead in the future. 

What we have succeeded in doing is to show that the quantum hypothesis is 
experimentally falsifiable. We have made the first step in the phenomenology of the 
quantum brain and this has opened up the way for more experiments and will hopefully 
make this hypothesis more attractive to mainstream theoretical and experimental 
physicists and biologists. 

It certainly is an astonishing premise that neurobiological processes taking place 
in a fly's brain are fundamentally tied to quantum events and this brings us full circle to 
the long conjectured connection between quantum physics and human consciousness.

\section{Acknowledgments}
The work of DVN supported in part by grant no. DE--FG--0395ER40917.
\newpage
%======================== REFERENCES =====================================
%========================================================================
%          MACROS FOR REFERENCES
%========================================================================
%=========================================================================
%atbib
               
%\bigskip
%\medskip

\bibliographystyle{unsrt}

\begin{thebibliography}{99}

\bibitem{pen1}Penrose, R. "The Emperor's New Mind". (Oxford University Press, Oxford 1989)
\bibitem{pen2}Penrose, R. "Shadows of the Mind". (Oxford University Press, Oxford 1994)
\bibitem{ham1}Hameroff, S.R. \& Penrose R. in Toward a Science of consciousness: the first Tucson discussion and debates. Edited by Hameroff S. R., Kaszniak, A.W. \& Scott. 
A.C.  (MIT Press, Cambridge), 507-539 (1996)
\bibitem{nan1}	Nanopoulos D.V. hep-ph/9505374 from XV Brazilian National Meeting on Particles and Fields, (1994) and Physics Without Frontiers Four Seas Conference (1995) 
\bibitem{mn1}	Mavromatos, N.E. \& Nanopoulos, D.V., in Advances in Structural Biology 5 283-318 (JAI Press Inc, London 1998) Malhotra S.K., Tuszynski J.A. Eds. Also in quant-ph/9802063
\bibitem{sat1}	Sataric M.V., Tuszynski J.A. \& Zakula R.B. Phys. Rev. E48 589 (1993)
\bibitem{fro1}	Fr\" olich, H. Int J. Quantum Chem. 2, 641-649 (1968)
\bibitem{fro2}	Fr\" olich, H. Nature 316, 349-351 (1970)
\bibitem{fro3}	Fr\" olich H.\& Kremer F. Coherent excitations in biological systems (Springer-Verlag, 
New York) (1983) 
\bibitem{mn2}	Mavromatos, N.E. \& Nanopoulos, D.V. Int. J. Mod. Phys. B11 ****** (1997)
\bibitem{mn3}	Mavromatos, N.E. \& Nanopoulos, D.V. Int. J. Mod. Phys. B12 517-542 (1998)
\bibitem{teg1}	Tegmark, M. quant-ph/9507009 submitted to Phys.Rev. E (1995)
\bibitem{edel1}	Edelman, G., Gall, W.E. \& Cowan W.M. Synaptic Function (John Wiley \& Sons, New York) (1987)
\bibitem{vor1}	Voronin L.V. Synaptic modifications \& Memory – An electrophysiological Analysis (Springer-Verlag, New York) (1993) 
\bibitem{haas1}	Haas H.L. \& Buzsaki G. Synaptic Plasticity in the Hippocampus,  (Springer-Verlag, New York)  (1988) 
\bibitem{der1}	Dermietzel, R. Brain Research Reviews xxx C 97000 123-133 (1997)
\bibitem{fit1}	Fitzgerald, R. Physics Today 17-19 March (1999)
\bibitem{ven1}Ventura D. \& Martinez T. quant-ph/9807059 submitted to IEEE Transactions on 
Neural Networks (1998)
\bibitem{ham2}	Hameroff, S. R. \& Penrose, R. Scale in conscious experience: Is the brain too 
important to be left to specialists to study? 243-274 (Mahwah, Erlbaum) (1995)
\bibitem{vog1}Vogel, G. Science, 280, 5 June  (1998)
\bibitem{arr1}	Arriagada P. et. al. Neurology, 42 631 (1992)
\bibitem{fra1}	Franks,N.P. \& Lieb W.R. Nature 300 487-493 (1982)
\bibitem{fra2}	Franks,N.P. \& Lieb W.R. Nature 316 349-351 (1985)
\bibitem{ham3}	Hameroff, S.R. \& Watt, R.C. Anesth..Analg. 62 936-940 (1983)
\bibitem{kor1}	Koruga, D. L. Ann. NY Acad. Sci. 466, 953-955 (1986)
\bibitem{cla1}	Clark, I. J.Biochemistry and Bioenergetics 41 59-61 (1996)
\bibitem{mar1}	Martin, K.C., Casadio, A., Zhu, H., Yaping, E., Rose, J.C., Chen, M., Bailey, C.H. 
\& Kandel, E.R., Cell 91, 927-938, Dec. (1997). 
\bibitem{woo1}	Woolf, N.J. Neurobiology of Learning and Memory 66 258-266 (1996)
\bibitem{woo2}	Woolf, N.J. Progress in Neurobiology 55, 59-77 (1998)
\bibitem{woo3}Woolf, N.J., Zinnerman, M.D. \& Johnson, G.V. Brain Res. 821(1), 241-249 (1999)
\bibitem{mn4}	Mavromatos N.E., Nanopoulos D.V., Samaras I. \& Zioutas K. in Adnances in 
Structural Biology 5 127-134 (JAI Press Inc., London 1998)
\bibitem{kan1}	Kandel, E.R., Schwartz J.H., Jessel T.M. Essentials of neural science and behavior (Appleton \& Lange, Norwalk, Connecticut 1995)
\bibitem{fuk1}	Fukunaga, K., Muller, D. \& Miyamoto, E. Neurochem. Int. 28, 343-358 (1996)
\bibitem{sac1}	Sackett D., based on a presentation at the workshop Biophysics of the Cytoskeleton. Banff, August 18-22 1997, Canada
\bibitem{del1}	Del Giuduce, E., Preparata, G., Vitiello, G. Phys. Rev. Lett. 61 1085 (1988)
\bibitem{del2}	Del Giuduce, E, Doglia, S., Milani, M., Vitiello, G. Nucl. Phys. B251(FS13) 376 (1985 ibid B275 (FS 17) 185 (1986)
\bibitem{jibu1}	Jibu, M., Hagan, S., Hameroff, S.R., Pronbram, K., Yasue, K. Biosystems 32 195(1994)
\bibitem{ham4}	Hameroff, S., R. Am.J.Clin.Med.  2 149 (1978)
\bibitem{nog1}	Nogales, E., Wolf, G. \& Downing, G., Nature 39, 124-134 (1998)
\bibitem{shor1}	Shor, P. SIAM Journal of Computing, 26 no. 5 1474-1483 (1997)
\bibitem{gro1}	Grover, L. Proceedings of the 28th Annual ACM Symposium on the Theory of 
Computing (ACM, New York) 212-219 (1996)
\bibitem{sha1}	Sharf, Y. \& Cory, D. G. quant-ph/0004030 (2000)
\bibitem{pre1}	Preskill., J. Physics Today June (1999)
\bibitem{ste1}	Steane, A.M. Phys. Rev. Lett.,77 793 (1996)
\bibitem{sak1}	Sakurai, J. J. Modern Quantum Mechanics Revised Edition (Addison-Wesley, New York 1995)
\bibitem{dav1}	Davis, R.L., Neuron, 11 1-14 (1993)
\bibitem{dav2}	Davis, R.L., Physiological Reviews, 76(2)  299-317; (1996)
\bibitem{skou1}	Skoulakis, E.M.C. \& Davis, R.L. Neuron, 12 931-944 Nov. (1996)
\bibitem{tul1}	Tully T. \& Quinn, W. J. Comp.Physiol. 157, 263-277 (1985)
\bibitem{hir1}	Hirokawa, N.  The Neuronal  Cytoskeleton  Burgoyne R.D. (Wiley-Liss, New 
York.) 5-74 (1991)
\bibitem{amos1}	Amos, L.A. \& Klug, A. J. Cell Sci. 14 523-550. (1974)





%****************************************************************************
\end{thebibliography}

\hfill\vfill\eject
\end{document}